\documentstyle[aps,12pt,tighten,epsf,cite]{revtex}
\title{\vspace*{2.0in} A Mass Hierarchy from
Recoiling
D branes }
\vspace{1.0in}
\author{ {\bf G.K. Leontaris}$^{a,b}$ and
{\bf N.E.~Mavromatos}$^{c}$}
\vspace{1.0in}

\address{$^a$Theoretical Physics Division, Ioannina
University, GR-45110 Ioannina, Greece.\\
$^b$ CERN, Theory Division, 1211 Geneva 23, Switzerland. \\
$^{\,c}$Theoretical Physics Group, Department of Physics,
King's College London, Strand, London WC2R 2LS, U.K. \\}

\newcommand{\be}{\begin{equation}}
\newcommand{\ee}{\end{equation}}
\newcommand{\bea}{\begin{eqnarray}}
\newcommand{\eea}{\end{eqnarray}}

\begin{document}

\maketitle

\vspace*{2cm}
\begin{centering}

{\bf Abstract}

\end{centering}
\begin{abstract}

Using conformal field theory methods we construct a metric that
describes the distortion of space-time surrounding a
D(irichlet)-brane (solitonic) defect after being struck by another
$D$-brane. By
viewing our four-dimensional universe as such a struck brane,
embedded in a five-dimensional space-time, we argue
on the appearance of a
band of massive Kaluza-Klein excitations for the bulk graviton
which is localized in a region of the fifth dimension determined
by the inverse size of the band. The band incorporates the
massless mode (ordinary graviton) and its thickness is determined
essentially by the width of the Gaussian distribution describing
the (target-space) quantum fluctuations of the intersecting-brane
configuration.
\end{abstract}

\vspace*{-6in}
\begin{flushright}
CERN.TH/99-404 \\
hep--th/9912230
\end{flushright}
\vspace*{6in}

\newpage
\vskip0.3cm

\section{Introduction}

Considerable scientific interest has been concentrated recently on
the revival and extension   of the rather old idea that space-time
is actually
$(4+n)$-dimensional, with our four-dimensional world being a membrane
(Dirichlet  brane~\cite{SR,Polch})  of some
 string theory living in a $4+n$-dimensional bulk 
 space-time~\cite{rubakov86,antoniadis90,dimopo98,antoniadis98,
antoniadis99,randal99}.

In some of these models, the extra (bulk) dimensions are taken to
be relatively large, compared to the traditional Planck scale,
implying, for instance, a bulk gravitational scale at the range of
a few TeV~\cite{dimopo98,antoniadis98,antoniadis99}. Considerable
effort has been devoted to a discussion of possible
phenomenological consequences of these scenarios  in
immediate-future accelerators such as LHC.

In the case of extra compact dimensions, which is the one assumed
in \cite{dimopo98,antoniadis98,antoniadis99}, there are induced
modifications of the four-dimensional Newton's law, which may
become phenomenologically important for TeV scale
gravity~\cite{kehagias99,floratos99}. Notice that such
modifications are distinct from earlier modifications proposed in
the context of supergravity~\cite{zachos}. It is straightforward
to check that at least two extra dimensions are needed in order to
avoid contradiction with the known laws
of gravity at large (solar) distances. On the other hand, for
$n=2$, astrophysical considerations~\cite{Astro} imply a scale
$M\ge 10 $TeV which marginally solves the hierarchy problem. For
larger $n$ there are less restrictions, however, it has been argued
that in this approach the hierarchy
problem essentially is re-formulated in terms of another
parameter which is now the compactification volume.

In the above scenario, the experimental
success\cite{krause} of the inverse-square law of
Newton seemed to imply precisely four non-compact
dimensions only. More recently,
however, the work of ref.~\cite{randal99}
has demonstrated that the
situation is completely different in cases  where the higher-dimensional
metric
was not factorizable~\cite{Gogberashvili:1998vx}, namely the case
where there is
    a {\it warp} factor in front of
the  four-dimensional metric which depends on the  coordinates of
the  bulk extra dimensions. According to this approach, our
universe is a static flat domain-wall which, in the simplest case
of five dimensions, separates two regions of five-dimensional Anti
de Sitter (AdS) space-time. In its simplest
version~\cite{randal99}, the scenario is realized by introducing a
positive energy brane at the origin and a negative energy brane at
distance $z$ where our world is located  and where the graviton
amplitude is exponentially suppressed. Modifications to the above
picture with positive energy branes allowing also the possibility
of infinite extra dimensions, multi-brane solutions, and
supergravity embedded versions were considered in the
literature~\cite{Kehagias:1999ju,GW,Lyk,Csaki:1999jh,Oda,Brandhuber:1999hb,DHR,PK}.
Thus, it is worth noticing that the bulk dimensions are not
necessarily compact. The rather  important point of
\cite{randal99}, however, was the demonstration of the
localization of the bulk gravitational fluctuations on the
three-dimensional brane,  which plays the r\^ole of our world.
This localization property was demonstrated by mapping the problem
of the dynamics of these fluctuations into a one-dimensional
Schr\"odinger eigenvalue problem.

A characteristic feature of such models was the presence of a
massless mode for the
 graviton (in agreement with  Lorentz covariance on the brane)
together with a  continuum of  massive Kaluza-Klein (KK)
states on
the four-dimensional world.
These KK modes have
different properties as compared with the factorizable case.
The presence of such KK states
leads to corrections of the
four-dimensional Newton's law; such corrections, however, are
suppressed by quadratic powers of the inverse Planck mass scale, and
hence are unobservable for all practical purposes. In some variants of
the model~\cite{nam99} one considers a periodic lattice of three
branes, which generates bands in the Kaluza-Klein spectrum,
separated from the massless graviton mode by a gap.

As a result of the above localization, a solution to the mass
hierarchy emerges in the sense that the weak scale is
generated from a large scale of the order of Planck mass through
an exponential hierarchy, induced by the presence of the warp
factor in the metric of the four-dimensional world.

The above models are very attractive, and indeed may offer a
viable solution to the hierarchy problem. However, we find it
rather restrictive that the discussions so far were concetrated
only on static brane configurations without including dynamics.

Indeed, it is known~\cite{kogan96,mavro+szabo} that when one
considers scattering of strings (or branes) off a $D$-brane, there
is a non-trivial recoil of the latter which distorts the
surrounding space-time~\cite{kanti98}, implying a sort of back
reaction. Such a back reaction curves the space-time around the
stringy defect in an non-trivial way. What we shall argue in this
article is that, as a result of such a back reaction, one can
obtain a different sort of mass hierarchy from that of
\cite{randal99}, though the concept of an induced non-factorizable
bulk metric also appears here.

The recoil problem is treated at present perturbatively for heavy
branes, within the  context of a world-sheet logarithmic conformal
field theory~\cite{kogan96,mavro+szabo,lcft}. What we shall do in
this work is to construct explicitly the space-time deformation
due to the recoil of a 4-brane, viewed as our {\it Euclideanized }
four-dimensional space-time embedded in a higher (five)
dimensional bulk space-time, after being struck by another brane.
We shall demonstrate the localization  of a {\it thin band} of
 KK massive bulk graviton modes (including the massless one) on our
 four-dimensional world,  with thickness determined by a weak  supersymmetry
  breaking scale $\alpha$ due to recoil~\cite{adrian+mavro99}. We shall also
  demonstrate
the formation of an horizon at distances given by the inverse of
the thickness
 of the  band of the localized KK modes.
On this horizon there is localization of
the rest of the massive KK modes, with masses higher than $\sim
\alpha $. We shall also demonstrate that in this scenario the
induced modifications of the four-dimensional Newton's law are
suppressed by powers of
$\alpha/M_s^2$, where $M_s$ is the string scale which in our case
may be taken  to be close to the Planck scale $M_s \sim 10^{18}$
GeV. Hence, such corrections are  essentially unobservable for
$\alpha \sim $ TeV, which is the case dictated by the gauge
hierarchy in our universe, given that $\alpha$ is the scale of the
induced supersymmetry breaking  on the 4-brane.

The structure of the article is the following: In section 2 we
present the salient
 features of the world-sheet approach to the $D$-brane struck by another
$D$-brane  or string. In section 3, we construct the space-time
deformation due to the recoil effects and show that a
non-factorizable five-dimensional metric arises. We show the existence of a
horizon located at a distance $z=1/\alpha$ and discuss
 analytic continuation beyond the horizon.   In section 4
we show that this metric
 is a solution to the Einstein equations describing an AdS
universe with negative
  bulk-cosmological constant which vanishes at $z=0$. We further
show in the same section that  the linearized Einstein equation
leads to a
 Schr\"odinger type equation with attractive potential for graviton modes in
a thin band of mass up to order $m \le \sqrt{2}\alpha$, including
the massless graviton mode (expected on account of Lorentz
covariance on the observable brane world).
We associate the scale $\alpha$ with that of supersymmetry-breaking
on the 4-brane, as a result of the recoil process~\cite{adrian+mavro99},
and demonstrate that the corrections to
the four-dimensional Newton's law are suppressed by powers
of $\alpha/M_s^2$, with $M_s \sim 10^{18}$ GeV in our scenario.
Conclusions and outlook are presented in section 5.

\section{World-sheet approach to $D$-brane/$D$-brane scattering: a review}

We first review the world-sheet formalism based on logarithmic
operators that was developed in a series of
papers~\cite{kogan96,mavro+szabo,ellis96,ellis98}, for the
mathematical description of the recoil of a
$D$-brane when struck by a closed-string state or by another $D$-brane.
 Logarithmic conformal field theory~\cite{lcft}
lies on the border between finite conformal field theories and
general renormalizable two--dimensional quantum field theories. It
is the relevant tool~\cite{kogan96,mavro+szabo,ellis96} for this
problem, because the recoil process involves a change of state
(transition) in the string background, and as such is not
described by a conformal field theory. This change of state
induced by the recoil process can be described as a change in the
$\sigma$-model background, and as such is a non-equilibrium process.
This is reflected~\cite{ellis96,mavro+szabo} in the logarithmic
operator algebra itself.

As discussed in references~\cite{kogan96,ellis96,mavro+szabo} in
the case of $D$-brane string solitons, their recoil after
interaction with a closed-string (graviton) state is characterized
by a $\sigma$-model deformed by a pair of logarithmic
operators~\cite{lcft}:
\begin{equation}
C^I_\epsilon = \epsilon \Theta_\epsilon (X^I),\qquad
D^I_\epsilon = X^I \Theta_\epsilon (X^I), \qquad I \in \{0,\dots, 3\}
\label{logpair}
\end{equation}
defined on the boundary $\partial \Sigma$ of the string world
sheet. Here $X^I, I\in \{0, \dots, 3\}$ obey Neumann boundary
conditions on the string world sheet, and denote the brane
coordinates, whilst $\Theta_{\epsilon}(X^{I})$ is the
regularized step-function, to be defined below.
The remaining $y^i, i\in \{4, \dots, 9\}$ denote the
transverse bulk directions.

In the case of $D$-particles, which were examined
in~\cite{kogan96,ellis96,mavro+szabo}, the index $I$ takes the
value $0$ only, in which case the operators (\ref{logpair}) act as
deformations of the conformal field theory on the world-sheet. The
operator $U_i \int _{\partial \Sigma} \partial_n X^i D_\epsilon $
describes the movement of the $D$-brane induced by the scattering,
where $U_i$ is its recoil velocity, and $Y_i \int _{\partial
\Sigma} \partial_n X^i C_\epsilon $ describes quantum fluctuations
in the initial position $Y_i$ of the $D$-particle. It has been
shown rigorously~\cite{mavro+szabo} that the logarithmic conformal
algebra ensures energy--momentum  conservation during the recoil
process: $U_i = \ell_s g_s ( k^1_i + k^2_i)$, where $k^1 (k^2)$ is
the momentum of the propagating closed string state before (after)
the recoil, and $g_s$ is the string coupling, which is assumed
here to be weak enough to ensure that $D$-branes are very massive,
with mass $M_D=1/(\ell _s g_s)$, where $\ell _s$ is the string
length.

\begin{figure}[htb]
\begin{center}
\epsfxsize=4in
\bigskip
\centerline{\epsffile{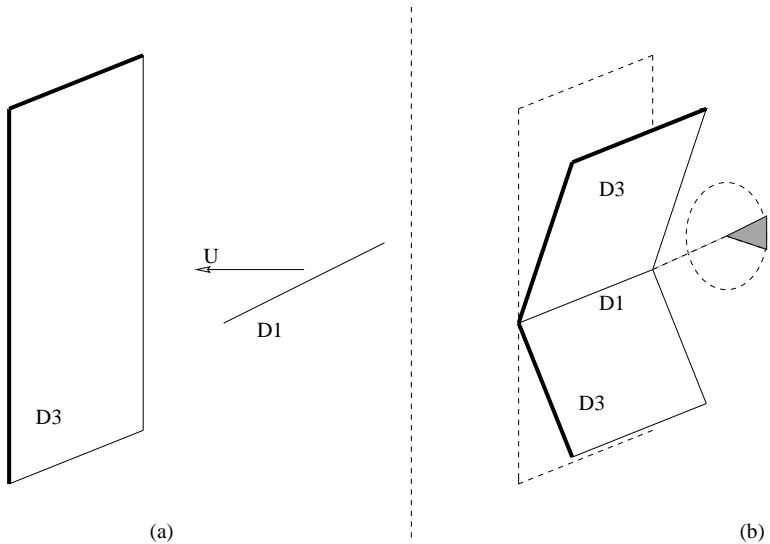}}
\vspace{0.2in}
\caption{\it Schematic representation of the folding effect
in $D$-brane/$D$-brane collisions:
(a) a $D1$ brane
moving with velocity $U$ along a `bulk' direction
perpendicular to a
$D3$ brane embedded in a
$D$-dimensional Euclidean
space-time E$_{D}$ strikes the $D3$ brane (b), which is then
folded, and the space-time around it is distorted into AdS$_3
\otimes {E}_{D-3}$. The dashed circle around the $D1$ direction in
(b) indicates the angular deficit that appears when the bulk
direction along which the $D1$ brane was moving is compactified to
a circle. A generalization to a higher-dimensional case for the
incident brane is straightforward. In that case the deficit (in
the compact case) is a higher-dimensional solid hyperangle.
\label{fig1}}
\end{center}
\end{figure}

In the case of $Dp$-branes, the pertinent deformations are
slightly more complicated. As discussed in~\cite{kogan96}, the
deformations are given by
\begin{equation}
\sum_{I} g^D_{Ii} \int _{\partial \Sigma}
\partial_n X^i D_\epsilon ^I \qquad{\mathrm and}\qquad\sum_{I}
g^C_{Ii} \int _{\partial \Sigma} \partial_n X^i C^I_\epsilon~.
\label{logpair2}
\end{equation}
The
$0i$ components of the two-index couplings
$g^{\alpha}_{Ii},~\alpha\in\{C,D\}$ include the collective momenta and
coordinates of the $D$-brane as in the $D$-particle case above,
but now there are additional couplings $g^{\alpha}_{Ii},~I\ne0$,
which describe the folding of the $D$-brane. Such a folding may be
caused by scattering with another macroscopic object, namely
another $D$-brane, propagating in a transverse direction, as shown
schematically in Fig.~\ref{fig1} for the case of a $D1$-brane
hitting a $D3$-brane. This situation is the most interesting to
us, since it generates an AdS$_3$ space, as we show below. For
symmetry reasons, in the situation depicted in Fig. \ref{fig1},
the folding of the $D3$-brane occurs symmetrically around the axis
of the $D1$-brane. In this case, the precise logarithmic operator
deformations shown in (\ref{logpair2}), which pertain only to the
spatial region $y_i > 0$ for the Dirichlet coordinates, should be
supplemented with their counterparts for the $y_i < 0$ region as
well. This would, in principle, require additional
$\Theta (\pm y_i)$
factors, which would complicate the analysis without
introducing any new points of principle. Therefore, for
simplicity, we restrict ourselves here to the $y_i > 0$  patch of
space-time, away from the hypersurface $y_i=0$. This will be
implicit in what follows.

The correct specification of the logarithmic pair in equation
(\ref{logpair2}) entails a regulating parameter
$\epsilon\rightarrow0^+$, which appears inside the $\Theta_\epsilon
(t)$ operator:
\be
\Theta_\epsilon (X^I) = \int \frac{d\omega}{2\pi}\frac{1}{\omega
-i\epsilon} e^{i\omega X^I} .
\label{thetaope}
\ee
In order to realize the logarithmic
algebra between the operators $C$ and $D$, one takes~\cite{kogan96}:
\begin{equation}
\epsilon^{-2} \sim \ln [L/a] \equiv \Lambda,
\label{defeps}
\end{equation}
where $L$ ($a$) are infrared (ultraviolet) world--sheet cutoffs.
The recoil operators (\ref{logpair2}) are
relevant, in the sense of the renormalization group for the
world--sheet field theory, having small conformal dimensions
$\Delta _\epsilon = -\epsilon^2/2$. Thus the $\sigma$-model
perturbed by these operators is not conformal for $\epsilon \ne
0$, and the theory requires Liouville
dressing~\cite{david88,distler89,ellis96}. Momentum conservation
is assured during the scattering process~\cite{mavro+szabo}.

The folding couplings $g_{Ii}^D \equiv g_{Ii},~I\in\{0, \dots,
p\}, ~i\in\{p+1, \dots, 9\}$, are relevant couplings with
world-sheet renormalization-group $\beta$ functions of the form
\begin{equation}
  \beta_{g_{Ii}} = \frac{d}{d t} g_{Ii} = -\frac{1}{2t} g_{Ii} , \qquad
t \sim \epsilon ^{-2}~.
\label{betaf}
\end{equation}
This implies that one may construct an exactly marginal set of
couplings ${\overline g_{Ii}}$ by redefining
\begin{equation}
{\overline g_{Ii}} \equiv \frac{g_{Ii}}{\epsilon}~.
\label{marginal}
\end{equation}
The renormalized couplings ${\overline g_{0i}}$ were shown
in~\cite{mavro+szabo} to play the r\^ole of the physical recoil
velocity of the $D$-brane, while the remaining ${\overline
g_{Ii}},~I\ne 0$, describe the folding of the $Dp$-brane for
$p\ne 0$. Here we shall assume, generalizing the case of
ref.~\cite{mavro+szabo} that the (bare) recoil couplings for all
$I$ are of equal strength and related to the transverse momentum
transfer as
\begin{equation}
  g_{Ii} =g_s \frac{\Delta P_i}{M_s}~, I=0, \dots ,m, ~ i=m+1, \dots D
\label{momtransf}
\end{equation}
for a D--brane embedded in a D-dimensional (bulk) space-time.

A technical but important remark is now in order, concerning
the world-sheet recoil formalism~\cite{kogan96}.
For reasons of convergence of the world-sheet
path integral, the Neumann coordinate $X^0$ must be
Euclideanized. It is only in this case that
the identification (\ref{defeps}), with $\epsilon^2 >0$,
leads to a mathematically
consistent logarithmic algebra of operators.
This can be understood simply by the fact that
in the pertinent world-sheet computations
of correlation functions of logarithmic operators
(\ref{logpair})
one encounters~\cite{kogan96}, due to (\ref{thetaope}),
the free propagator of the Neumann coordinates $X^I$:
\be
{\cal G}_0 = {\rm \lim}_{\sigma \rightarrow 0}<X^I(\sigma) X^J
(0)>_* \sim \eta_{IJ} \ln [L/a] \label{freeprop}
\ee
where $< \cdots >_*$ denotes world-sheet partition function with
respect to the free-string world-sheet action on a flat target
space-time manifold $\{ X^I \}$,  and $\eta^{IJ}$ is the target
space metric. For Euclidean world-sheets one takes
$\eta^{IJ}=\delta^{IJ}$, and this is essential for the convergence
of world-sheet path integral expressions entering in the
respective correlators. Indeed, let us illustrate this by a simple
example of the one-point function $<C>$. This involves (c.f.
(\ref{thetaope})) the computation of
$$
<\int _{-\infty}^{+\infty}\frac{d\omega}{\omega-i\epsilon}{\rm
\exp} \left(-i\omega X^0\right)>_* \sim \int
_{-\infty}^{+\infty}\frac{d\omega}{\omega-i\epsilon}{\rm \exp}
\left(-\frac{\omega^2}{2}<X^0X^0>_*\right).$$ There are
world-sheet ultraviolet infinities coming from the coincidence
limit of the $X^0$ propagator in this expression, which after
regularization give~\cite{kogan96}:
$$\int _{-\infty}^{+\infty}\frac{d\omega}
{\omega-i\epsilon}{\rm exp}\left(-\frac{\omega^2}{2}\eta^{00}\ln
[L/a]\right).$$ Such integrals are convergent only for Euclidean
$X^0$, which we have assumed in \cite{kogan96} and here.

The Euclideanization of the Neumann coordinates implies that in
our picture, of viewing the (3+1)-dimensional world as a brane,
the (longitudinal) Neumann coordinates will define a $D4$ domain
wall in the bulk space-time, which, after analytic continuation of
the coordinate $X^0$, will result in our four-dimensional
space-time. However, the analytic continuation will take place
only at the very end of the calculations. This will be very
important for our purposes here, and will always be understood in
what follows.

\section{Recoil-Induced Space-Time Metric Deformations}

As discussed in~\cite{ellis96,kanti98}, the deformations
(\ref{logpair}) create a local distortion of the space-time
surrounding the recoiling folded $D$-brane, which may be
determined using the method of Liouville dressing.
In~\cite{ellis96,kanti98} we concentrated on describing the
resulting space-time in the case when a $D$-particle, embedded in
a $D$-dimensional space-time, recoils after the scattering of a
closed string off the $D$-particle defect. To leading order in
the recoil velocity $u_i$ of the $D$-particle, the resulting
space-time was found, for times $t \gg 0$ long after the
scattering event at $t=0$, to be equivalent to a Rindler wedge,
with apparent `acceleration' $\epsilon u_i$~\cite{kanti98}, where
$\epsilon$ is defined above (\ref{defeps}).
For times $t < 0$, the space-time is flat Minkowski~\footnote{There is
hence a discontinuity at $t =0$, which leads to particle
production and decoherence for a low-energy spectator field theory
observer who performs local scattering experiments
long after the scattering, and far away from the
location of the collision of the closed string with the
$D$ particle~\cite{kanti98}.}.

This situation is easily generalized to $Dp$-branes~\cite{emw99}.
The folding/recoil deformations of the $Dp$-brane (\ref{logpair2})
are relevant deformations, with anomalous dimension
$-\epsilon^2/2 $, which disturbs the conformal invariance of the
$\sigma$ model, and restoration of conformal invariance requires
Liouville dressing~\cite{distler89}. To determine the effect of
such dressing on the space-time geometry, it is essential  to
write~\cite{ellis96} the boundary recoil deformations as  bulk
world-sheet deformations
\begin{equation}
\int _{\partial \Sigma} {\overline g}_{Iz} x\Theta_\epsilon (x)
\partial_n z =
\int _\Sigma \partial_\alpha \left({\overline g}_{Iz} x\Theta_\epsilon (x)
\partial ^\alpha z \right)
\label{a1}
\end{equation}
where the ${\overline g}_{Iz}$ denote the renormalized
folding/recoil couplings (\ref{marginal}), in the sense discussed
in~\cite{mavro+szabo}. As we have already mentioned, such
couplings are marginal on a flat world sheet.
The operators (\ref{a1}) are marginal also on a curved
world sheet, provided~\cite{distler89} one dresses the (bulk)
integrand by multiplying it by a factor $e^{\alpha_{Ii}\phi}$,
where $\phi$ is the Liouville field and $\alpha_{Ii}$ is the
gravitational conformal dimension, which is related to the
flat-world-sheet anomalous dimension $-\epsilon^2/2$ of the recoil
operator, viewed as a bulk world-sheet deformation, as
follows~\cite{distler89}:
\begin{equation}
\alpha_{Ii}=-\frac{Q_b}{2} +
\sqrt{\frac {Q_b^2}{4} + \frac {\epsilon^2}{2} }
\label{anom}
\end{equation}
where $Q_b$ is the central-charge deficit of the bulk world-sheet
theory. In the recoil problem at hand, as discussed
in~\cite{kanti98},
\be
Q_b^2 \sim \epsilon^4/g_s^2  > 0
\label{centralcharge}
\ee
for weak folding deformations $g_{Ii}$, and hence one is
confronted with a {\it supercritical} Liouville theory. This
implies a {\it Minkowskian signature} Liouville-field kinetic term
in the respective $\sigma$-model~\cite{aben89}, which prompts one
to interpret the Liouville field as a time-like target-space
field. However, in our context, this will be a {\it second} time
coordinate~\cite{emn98}, which is independent of the
(Euclideanized) $X^0$. The presence of this second `time' for us
will not affect the physical observables, which will be defined
for appropriate slices of fixed Liouville coordinate, e.g. $\phi
\rightarrow \infty$, or equivalently
$\epsilon \rightarrow 0$.
From the expression (\ref{centralcharge}) we conclude (c.f.
(\ref{anom})) that $\alpha_{Ii} \sim \epsilon $ to leading order
in perturbation theory in $\epsilon$, to which we restrict
ourselves here.

We next remark that, as the analysis of~\cite{ellis96} indicates,
the $X^I$-dependent field operators
$\Theta_\epsilon (X^I)$ scale as follows with $\epsilon$:
$\Theta_\epsilon(X^I) \sim e^{-\epsilon X^I}
\Theta(X^I)$, where $\Theta(X^I)$ is a Heavyside step function
without any field content, evaluated in the limit $\epsilon \rightarrow 0^+$.
The bulk deformations, therefore, yield the following
$\sigma$-model terms:
\begin{equation}
\frac{1}{4\pi \ell_s^2}~\int _\Sigma 
\sum_{I=0}^{3} \left( \epsilon^2 {\overline g}^C_{Ii} + \epsilon 
{\overline g}_{Ii} X^I\right)
e^{\epsilon(\phi_{(0)} - X^I_{(0)})}\Theta(X^I_{(0)})
\partial_\alpha \phi 
\partial^\alpha y_i~
\label{bulksigma}
\end{equation}
where the subscripts $(0)$ denote world-sheet zero modes, and 
${\overline g}^C_{0i}=y_i$.

Upon the interpretation of the Liouville zero mode $\phi_{(0)}$ as
a (second)
time-like coordinate, the deformations (\ref{bulksigma}) yield
space-time metric
deformations (of the generalized space-time
with two times).
The metric components
for fixed Liouville-time slices can be
interpreted in~\cite{ellis96}
as expressing the distortion of the space-time
surrounding the recoiling $D$-brane soliton.

For clarity,
we now drop the subscripts $(0)$ for the rest of this paper,
and we work in a region of space-time
on the $D3$ brane such that $\epsilon (\phi - X^I)$ is finite
in the limit $\epsilon \rightarrow 0^+$.
The resulting space-time distortion is therefore
described by the metric elements
\begin{eqnarray}
&~& G_{\phi\phi} = -1, \qquad G_{ij} =\delta_{ij}, \qquad
G_{IJ}=\delta_{IJ}, \qquad G_{iI}=0,   \nonumber \\
&~& G_{\phi i} = \left(\epsilon^2 {\overline g}^C_{Ii} +
 \epsilon {\overline g}_{Ii}X^I \right)\Theta (X^I)~,
\qquad i=4, \dots 9,~~I=0, \dots 3
\label{gemetric}
\end{eqnarray}
where the index $\phi $ denotes Liouville `time', not to be confused
with the Euclideanized time which is one of the $X^I$.
To leading order in $\epsilon {\overline g}_{Ii}$,
we may ignore the $\epsilon^2 {\overline g}^C_{Ii}$ term.
The presence of $\Theta(X^I)$ functions and
the fact that we are working in the region $y_i >0$
indicate that
the induced space-time is piecewise continuous~\footnote{The
important implications for non-thermal particle production
and decoherence for a spectator low-energy field theory
in such space-times were discussed in~\cite{kanti98,ellis96}, where only
the $D$-particle recoil case was considered.}.
In the general recoil/folding case considered in this article,
the form of the resulting patch of the surrounding
space-time can be determined fully if one computes
the associated curvature tensors, along the lines
of~\cite{kanti98}.

We now conclude this section with some
remarks about the metric (\ref{gemetric}).
First we restrict ourselves to the case
of a single Dirichlet dimension $z$,
playing the r\^ole of a bulk dimension
in a set up where there are $X^I$, $I=0,\dots3$
Neumann coordinates parametrizing a D4 (Euclidean) brane
(our four-dimensional space-time).
Upon performing the time transformation
$\phi \rightarrow \phi - \frac{1}{2}\epsilon {\overline g}_{Iz} X^I z $, the
line element of the above-mentioned space-time
becomes:
\begin{eqnarray}
&~&ds^2 =-d\phi^2 + \left(\delta_{IJ}
-\frac{1}{4}\epsilon^2{\overline g}_{Iz}{\overline g}_{Jz}~z^2\right)~dX^I dX^J
+ \left(1 + \frac{1}{4}\epsilon ^2
{\overline g}_{Iz}{\overline g}_{Jz}~X^I~X^J\right)~dz^2
- \epsilon {\overline g}_{Iz}~z~dX^I~d\phi~, \nonumber \\
\label{bendinglineel}
\end{eqnarray}
where $\phi$ is the Liouville field (which, we remind the reader,
has Minkowskian signature, in the case of supercritical
strings we are dealing with here.)

One may now invoke a general coordinate transformation on the
brane $X^I$ so as to diagonalize the pertinent induced-metric
elements in (\ref{bendinglineel})~\footnote{Note that general
coordinate invariance is assumed to be a good symmetry on the
brane, away from the `boundary' $X^I=0$.}. For instance, to
leading order in the deformation couplings ${\overline
g}_{Iz}{\overline g}_{Jz}$, one may redefine the $X^I$ coordinates
by \begin{eqnarray} X^I &\rightarrow& X^I -\frac{\epsilon^2}{8}z^2
{\overline g}_{Iz} \sum_{J \ne I}{\overline g}_{Jz}X^J,\nonumber\\
z &\rightarrow& z \left(1 + \frac{\epsilon^2}{8}\sum_{I \ne J}
{\overline g}_{Iz}{\overline g}_{Jz}X^{I}X^{J}\right)
\end{eqnarray}
 which leaves only the diagonal elements  of the
metric tensor on the (redefined) hyperplane $X^I$. In that case,
the metric becomes (to leading order in $g_{Iz}^2$):
\begin{eqnarray}
&~&ds^2 =-d\phi^2 + \left(1
-\alpha^2 ~z^2\right)~(dX^I)^2
+ \left(1 + \alpha^2 ~(X^I)^2\right)~dz^2 - \epsilon {\overline g}_{Iz}~z~dX^I~d\phi~,
\nonumber \\
&~& \alpha=\frac{1}{2}\epsilon {\overline g}_{Iz} \sim g_s |\Delta P_z|/M_s
\label{bendinglineel3}
\end{eqnarray}
where in the last expression
we wanted to make clear that, upon utilizing
(\ref{marginal}),(\ref{momtransf}), one can actually express the parameter
$\alpha$ (in the limit $\epsilon \rightarrow 0^+$)
in terms of the (recoil) momentum transfer along the bulk direction.
As we shall see later on, this parameter is responsible for the mass hierarchy
in the problem, assuming that the string scale $M_s$ is close to
Planck mass scale $10^{18}$ GeV, for ordinary string-theory
couplings of order $g_s^2/2\pi =1/20$.
The above metric element is derived in the case where $
\epsilon {\overline g}_{Iz}z << 1$.

A last comment concerns the case in which
the metric (\ref{bendinglineel3}) is {\it exact}, i.e. it holds
to all orders in ${\overline g}_{Iz}z$.
This is the case where
there is no world-sheet
tree level momentum transfer. This naively corresponds to the case
of static intersecting branes. However, the whole philosophy of
recoil~\cite{kogan96,mavro+szabo} implies that even in that case
there are quantum fluctuations induced by summing up genera on the
world-sheet. The latter implies the existence of a statistical
distribution of logarithmic deformation couplings of Gaussian type
about a mean field value ${\overline g}^{c}_{Iz}=0$. Physically,
the couplings
${\overline g}_{Iz}$ represent recoil velocities of the intersecting
branes,
hence the situation of a Gaussian fluctuation about a zero mean
value represents the effects of quantum fluctuations about the
zero recoil velocity case, which may be considered as a quantum
correction to the static intersecting brane case. Such Gaussian
quantum fluctuations arise quite naturally by summing up higher
world-sheet topologies~\cite{mavro+szabo}. We therefore consider
taking a statistical average
$<< \cdots >>$ of the line element (\ref{bendinglineel})
\begin{eqnarray}
&~&<<ds^2>> =-d\phi^2 + \left(1
-\frac{1}{4}\epsilon^2
<<{\overline g}_{Iz}{\overline g}_{Jz}>>~z^2\right)~dX^I dX^J
+ \nonumber \\
&~& \left(1 + \frac{1}{4}\epsilon ^2
 <<{\overline g}_{Iz}{\overline g}_{Jz}>>~X^I~X^J\right)~dz^2
- \epsilon <<{\overline g}_{Iz}>>~z~dX^I~d\phi~, \nonumber \\
\label{bendinglineel2}
\end{eqnarray}
where
\be
<< \cdots >>=\int _{-\infty}^{+\infty}d{\overline g}_{Iz}
\left(\sqrt{\pi}\Gamma \right)^{-1}
 e^{-{\overline g}_{Iz}^2/\Gamma^2} (\cdots) \label{gauss}
\ee
where the width $\Gamma$ has been calculated
in \cite{mavro+szabo}, after proper summation over world-sheet
genera, and in fact is found to be
proportional to the string coupling $g_s$.

Obviously, from (\ref{gauss}), and
assuming that $g_{Iz}=|U_i|$, where $U_i=g_s \Delta P_i/M_s$
is the recoil velocity~\cite{kogan96,mavro+szabo},
the average line element
$ds^2$ becomes:
\begin{eqnarray}
&~&<<ds^2>> =-d\phi^2 + \left(1
-\alpha^2 ~z^2\right)~(dX^I)^2
+ \left(1 + \alpha^2 ~(X^I)^2\right)~dz^2,
\nonumber \\
&~& \alpha=\frac{1}{2\sqrt{2}}\epsilon \Gamma 
\label{bendinglineel3ab}
\end{eqnarray}
The definition of $\alpha$ comes from evaluating the quantity
$<<{\overline g}_{Iz}^2>>$ using the statistical distribution (\ref{gauss}).
Thus, in that case, averaging over quantum fluctuations leads to a
metric of the form (\ref{bendinglineel3}), but with a parameter
$\alpha$ much smaller, being determined by the width (uncertainty)
of the pertinent quantum fluctuations~\cite{mavro+szabo}.
The metric (\ref{bendinglineel3ab})
is exact, in contrast to the metric (\ref{bendinglineel3})
which was derived for $z << 1/\alpha$. However, for our purposes
below we shall treat both
metrics as exact solutions of some string theory
associated with recoil.

An important feature of the line element (\ref{bendinglineel3ab}) is
the existence of an {\it horizon} at $z=1/\alpha$ for {\it Euclidean}
Neumann coordinates $X^I$. Also notice that the Liouville field
$\phi$ has decoupled, upon the averaging procedure, and this allows
one to consider slices of this field, defined by $\phi$=const, on
which  the physics of the observable world can be studied. From a
world-sheet renormalization-group view point this slicing
procedure corresponds to selecting a specific point in the
non-critical string theory space. Usually, the infrared fixed
point
$\phi \rightarrow \infty$ is selected. In that case, from (\ref{defeps}),
one considers a slice for which $\epsilon^2 \rightarrow 0$.
But any other choice could do, so $\alpha$ may be considered
a small but otherwise arbitrary parameter of our effective theory.

The presence of an horizon raises the issue of how one
could analytically continue so as to pass to the space beyond the horizon.
The simplest way, compatible, as we shall show later with the low-energy
Einstein's equations, is to take the absolute value of $1-\alpha^2 z^2$
in the metric element (\ref{bendinglineel3}).
We therefore consider the following metric defined in all space $z \in R$
at a slice of the Liouville time $\phi$=const:
\begin{eqnarray}
&~&ds^2_{f} = \left|1
-\alpha^2 ~z^2\right|~(dX^I)^2
+ \left(1 + \alpha^2 ~(X^I)^2\right)~dz^2,
%\nonumber
%\\
%&~& \alpha=\frac{1}{2}\epsilon \sqrt{\pi}\Gamma^3
\label{bendinglineel3b}
\end{eqnarray}
For small $\alpha$, which is the case studied here,
and for Euclidean
Neumann coordinates $X^I$, the scale factor in front of the
$dz^2$ term does not introduce any singular behaviour, and hence
for all qualitative purposes we may study the following metric element:
\begin{eqnarray}
&~&ds^2_{f} = \left|1
-\alpha^2 ~z^2\right|~(dX^I)^2
+ ~dz^2,
%\nonumber \\
%&~& \alpha=\frac{1}{2}\epsilon \sqrt{\pi}\Gamma^3
\label{bendinglineel4}
\end{eqnarray}
which is expected to share all the qualitative
features of the full metric (\ref{bendinglineel3b})
induced by the recoil process in the case of an uncompactified
`bulk' Dirichlet dimension $z$ we restrict ourselves
here~\footnote{For the case of compact dimension $z$
the situation changes drastically,
since in that case,
for compact $z$ and at fixed $X^I \sim 1/\epsilon \gg 0$ and $t \gg 0$,
such that $\alpha^2~X^2 \simeq g_{Xz}^2/4 $,
we observe from the metric (\ref{bendinglineel3b})
that there exists a deficit angle
in the circle around $z$~\cite{emw99}:
\begin{equation}
\delta \simeq (\pi g_{Iz}^2/4)
\label{deficit}
\end{equation}
implying the dynamical formation
of a conical-like singularity.
Such singularities in general break bulk
space-time supersymmetry~\cite{adrian+mavro99}.
However, in view of the fact that the
folded $D$- brane is an excited state of
the string/$D$-brane system, the phenomenon should be
viewed as a symmetry obstruction rather
than a spontaneous breaking of symmetry, in the sense
that, although the ground state of the
string/$D$-brane system is supersymmetric, recoil
produces a particular excited state
that does not respect that symmetry~\cite{witten95}.
We shall not, however, deal any further with the
compact case in what follows, but instead assume a non-compact
bulk dimension.}.

A point that we would like to make concerns the fact that,
formally, our analysis leading to (\ref{bendinglineel4}) is valid
in the region of bulk space-time for which $z > 0$. However, one
may consider a {\it mirror} extension of the space-time for the
region $z < 0$, which we assume in this article. {}From now on,
therefore, we treat the metric (\ref{bendinglineel4}) as being
defined over the entire real axis for the bulk coordinate $z \in
R$. However, to make contact with the original recoil picture we
restrict ourselves in regions of space-time for which
$X^I >0$.

\section{A Mass Hierarchy from Recoiling $D$-branes}

In this section we show that the metric obtained by the dynamical
mechanism of $D$-brane scattering predicts a natural scale
hierarchy. A crucial role is played by the value of the only
parameter of the theory, i.e. $\alpha$, which is directly related
to the $D$-brane recoil and appears in the  {\it warp} factor in
front of the four-dimensional part of the metric.  With the above
in mind we now write the metric (\ref{bendinglineel4}) as:
\begin{eqnarray}
&~&ds^2_{f} = e^{-2\sigma (z)}
~(dX^I)^2 + dz^2,
\nonumber \\
&~&\sigma (z) =-\frac 12 \ln\left( \left|1 -\alpha^2
~z^2\right|\right) \label{bulk}
\end{eqnarray}
The only non-zero components of the Christoffel symbol
corresponding to the metric (\ref{bulk}) read (in Euclidean
signature for $X^I, I=0,\dots 3$):
\bea
&~& \Gamma^{0}_{04} = \Gamma _{40}^{0} = \sigma '(z) \nonumber \\
&~& \Gamma^{4}_{00} = -\sigma '(z) e^{-2\sigma (z)} \nonumber \\
&~& \Gamma^{4}_{ii} = \sigma '(z) e^{-2\sigma (z)}~,
\qquad i=1,2,3 \nonumber \\
&~& \Gamma^{i}_{i4} = \Gamma _{4i}^{i} = -\sigma '(z)
\label{christof}
\eea
where the prime denotes differentiation with respect to $z$.
Notice that in the case of Minkowskian signature for the Neumann time
coordinate
$X^0$, the only change will be
$\Gamma^{4}_{00}
\rightarrow -\Gamma^{4}_{00}$. This implies a
similar sign change for the corresponding components
of the Ricci curvature $R_{00} \rightarrow -R_{00}$.
The curvature scalar therefore remains unchanged
upon the analytic continuation of the time variable $X^0$.

For future use we note the following mathematical identities:
\bea
&~& \sigma '(z) = -\frac{1}{2}\frac{\alpha }{|1 + \alpha z|}\left[
\Theta(1 + \alpha z) - \Theta(-1-\alpha z)\right]
+\frac{1}{2}\frac{\alpha }{|1 - \alpha z|}\left[
\Theta(1 - \alpha z) - \Theta(-1+\alpha z)\right]~, \nonumber \\
&~& \sigma ''(z) = \frac{1}{2}\frac{\alpha ^2}{(1 + \alpha z)^2}
\left[\Theta (1 + \alpha z) - \Theta(-1-\alpha z)\right]+
\frac{1}{2}\frac{\alpha ^2}{(1-\alpha z)^2}
\left[\Theta(1 - \alpha z) - \Theta(-1+\alpha z)\right] \nonumber \\
&~& -\frac{\alpha ^2}{|1 + \alpha z|}
\delta (1 + \alpha z)
-\frac{\alpha ^2}{|1 - \alpha z|}\delta (1 - \alpha z)~.
\label{mathident}
\eea

We next check on whether the metric (\ref{bulk})
is a solution of the Einstein's equations:
\bea
&~&    R_{\mu\nu}-\frac{1}{2}G_{\mu\nu}R = T_{\mu\nu}~, \nonumber \\
&~& T_{\mu\nu} = -\frac{1}{4M_s^3} G_{\mu\nu}\Lambda -
\frac{1}{4M_s^3}\sum_{i} \sqrt{G^{(i)}} G_{IJ}^{(i)} \delta^{I}_\mu
\delta^{J}_\nu
V_{(i)}(z)/\sqrt{G}~,
\mu, \nu=0, \dots 4~, I,J = 0, \dots 3,
\label{einstein}
\eea
where $M_s$ is the string mass scale and
$\Lambda$ is a cosmological constant in the bulk space-time,
and the sum $\sum_{i}$ is over possible $D$-brane defects.
The index (i) denotes quantities pertaining strictly to
such $D$-brane domain walls. In our case we can assume
$i=1$, since originally we
have a (struck) $D$-brane at the origin $z=0$.
Note that, in a similar spirit to \cite{randal99}, we have subtracted
a vacuum energy contribution, proportional to $V(z)$ from such $D$-brane
defects.

It is easy to check from (\ref{mathident})
that, by placing such domain walls at the horizon points
$z = \pm 1/\alpha$, one obtains that the metric (\ref{bulk})
is indeed a solution of (\ref{einstein}), provided
that
\bea
3\sigma '' (z) = \frac{1}{4M_s^3}\sum_{i} V_{(i)}(z)
\frac{\sqrt{G^{(i)}}G_{00}^{(i)}}{\sqrt{G}e^{-2\sigma}}~, \nonumber
&~& (\sigma ')^2 = -\frac{1}{24 M_s^3}\Lambda
\label{simpleqs}
\eea
For a single $D$-brane at $z=0$, the solution is:
\bea
&~& \frac{\Lambda}{24M_s^3}=-\frac{1}{4}\frac{\alpha ^2}{(1 + \alpha z)^2}
-\frac{1}{4}\frac{\alpha ^2}{(1 - \alpha z)^2}+\frac{1}{2}
\frac{\alpha ^2}{(1 - (\alpha z)^2)}{\cal E}, \nonumber \\
&~& \frac{V(z)}{4M_s^3}=\frac{3}{2}\frac{\alpha ^2}{(1 + \alpha z)^2}
+\frac{3}{2}\frac{\alpha ^2}{(1 - \alpha z)^2} - \nonumber \\
&~& \frac{3\alpha ^2}{|1 + \alpha z|}\delta (1 + \alpha z)
- \frac{3\alpha ^2}{|1 - \alpha z|}\delta (1 - \alpha z)
\label{explsol}
\eea
where ${\cal E}=
\left[\Theta(1 + \alpha z)-\Theta (-1-\alpha z)\right]
\left[\Theta(1 - \alpha z)-\Theta (-1+\alpha z)\right]=
+1$, if $-1/\alpha < z < 1/\alpha$, and ${\cal E}=-1$
otherwise.

The negative cosmological constant (anti-de-Sitter type Universe)
is a generic feature of intersecting branes~\cite{randal99}, but
also of the recoil formalism~\cite{emn98,emw99}, and signals
compatibility with space-time supersymmetry in the case of static
non-recoiling intersecting branes~\cite{randal99}. Notice,
however, that the cosmological constant {\it vanishes} on the
original brane $z=0$ and becomes infinitely negative on the
horizon $|z|=1/a$.

On the other hand,
the vacuum energy distribution on the brane hypersurface, $V_{(z)}$,
is positive ($V(0)=12\alpha ^2$) at $z=0$, which is compatible with the
fact that the recoil excites the $D$-brane at $z=0$, and the
excitation   energy is of the same order as the kinetic energy
transfer, due to energy-momentum conservation~\cite{mavro+szabo}.
$V_{(z)}$ also blows up negative at the horizon, signaling the formation
of domain walls there.

Next we consider the issue of localization of bulk graviton states
inside the horizon $-1/\alpha < z < 1/\alpha$.
To this end, we follow
\cite{randal99} and
use the following ansatz for separating variables
$X^I$ and $z$, as far as
(small) quantum fluctuations of
the bulk graviton state
${\hat h}(X^I,z)$ about the background (\ref{bulk})
are concerned:
\begin{equation}
 {\hat h}(X^I,z)=\lambda (z) e^{ip^E_IX^I}
\label{fluct}
\end{equation}
where the notation $p^E_I$ in the momenta on the brane has been
explicitly state to remind the reader that we are working on a
Euclidean set up for $\{ X^I \}$, and hence for massive
KK excitations, of mass squared $m^2>0$ the on-shell
condition should read
\begin{equation}
(p^E_I)^2=-m^2 < 0
\label{onshell}
\end{equation}
The equation for such small fluctuations, can be obtained
by linearizing  Einstein's equations (\ref{einstein}) around the
AdS background and choosing appropriate gauge for the fluctuations
of the metric.  The final equation then reads:
\begin{equation}
\left(   -\partial_I \partial^I - \partial_z \partial^z +
{\cal V}(z)
\right){\hat h}(X^I, z)=0
\label{eqfluct}
\end{equation}
where the ``potential'' ${\cal V}(z)$ arises from curvature.

\begin{figure}[htb]
\begin{center}
\epsfxsize=4in
\bigskip
\centerline{\epsffile{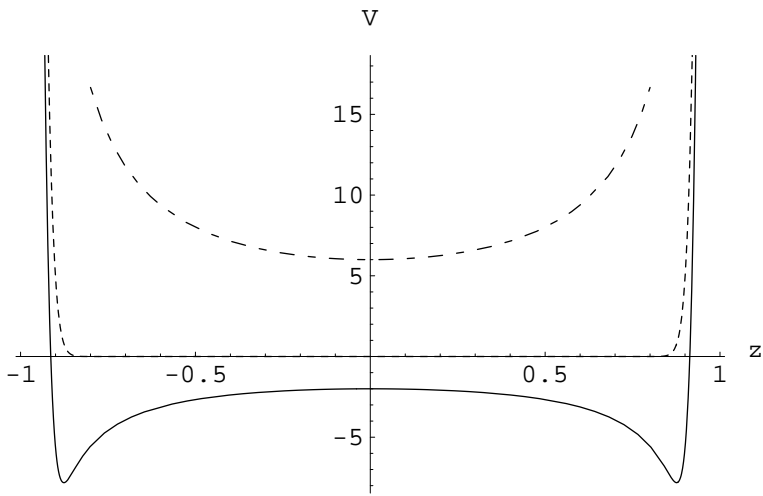}} \vspace{0.2in} \caption{\it
Schematic representation of the potential of the equivalent
Schr\"odinger equation in the bulk direction. The solid curve
corresponds to the massless case while the dashed one represents
the potential for the critical value $m=\sqrt{2} \alpha$. Beyond
this value, the massive  gravitons wavefunctions are no-longer
localized on the $z=0$ $D3$-brane; the dashed-dotted curve
represents the potential for such a case. \label{fig2}}
\end{center}
\end{figure}

Upon introducing the ansatz (\ref{fluct}), and using
(\ref{onshell}), the above equation becomes a one-dimensional
Schr\"odinger-type eigenvalue equation
for the bulk modes $\lambda (z)$~\cite{randal99}
\begin{equation}
-\lambda '' (z) + \left(4 (\sigma ')^2 - 2 \sigma '' \right)\lambda (z) =
-m^2 e^{2\sigma}\lambda (z)
\label{lambdaeq}
\end{equation}
where the various quantities are given in (\ref{mathident})
for the problem at hand.
It is important to note the {\it minus}
sign in front of the mass term on the right-hand-side
of (\ref{lambdaeq}), which is due to the
Euclidean nature
of $X^I$ hyperplane (\ref{onshell}).
As we shall see soon this will play an important
physical r\^ole.
Substituting (\ref{mathident}) in (\ref{lambdaeq})
one obtains after some straightforward algebra:
\begin{eqnarray}
&~&-\lambda '' (z) - \left(\frac{2\alpha ^2}{|1 - \alpha^2 z^2 |}{\cal E}
-  \frac{\alpha ^2}{|1 + \alpha z |}\delta (1 + \alpha z)
 - \frac{\alpha ^2}{|1 - \alpha z |} \delta (1-\alpha z)+
\frac{m^2}{|1 - \alpha^2 z^2 |}\right)\lambda (z) =0~, \nonumber \\
&~& {\cal E}=+1 \qquad -1/\alpha < z < 1/\alpha~,
\qquad {\cal E}=-1~~~ {\rm otherwise}
\label{theequation}
\end{eqnarray}
The equation (\ref{lambdaeq}) has the form of a one-dimensional
Schr\"odinger's equation.
The potential is drawn in figure \ref{fig2} for various values of
the mass parameter $m$.

\begin{figure}[htb]
\begin{center}
\epsfxsize=4in
\bigskip
\centerline{\epsffile{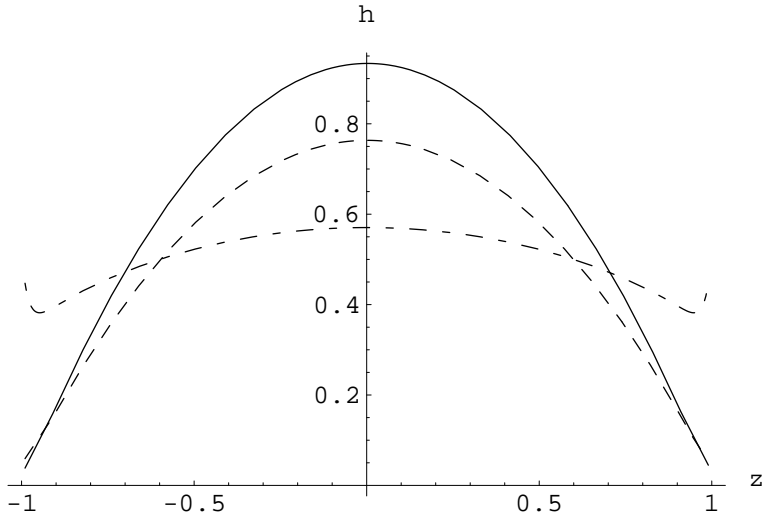}}
\vspace{0.2in}
\caption{\it The solution of the Schr\"odinger equation
for three Kaluza-Klein modes with masses
 $0\le m< \sqrt{2}\alpha$. Solid curve corresponds to $m=0$
wavefunction.
We observe that the modes  are localized within the horizon
$|z| < 1/\alpha$, while they start to spread as $m$ grows.
For $m\ge \sqrt{2}\alpha$ they are peaked on the boundaries
$\pm 1/\alpha$.
\label{fig3}}
\end{center}
\end{figure}

We observe the following:

\begin{itemize}
\item{} For mass parameters $0 < m < m_{cr}\equiv \sqrt{2}\alpha $ the potential
is attractive, and the wave function peaks at $z=0$ (see figure \ref{fig3}).

\item{} For mass parameter $m > \sqrt{2}\alpha$ the wavefunctions
are sharply peaked at the horizon $|z|=1/\alpha$ .

\end{itemize}

{}For illustration purposes, the wavefunctions (arbitrarily normalized) 
for three distinct cases, $m=0, m=0.2 m_{cr}$ and $m = 0.8 m_{cr}$
 are depicted in figure~(\ref{fig3}).
{}For $m=0$ (massless graviton state), the corresponding wavefunction (solid curve)
peaks at the $D4$-brane at $z=0$. 
 As $m$ approaches the critical value $m_{cr}$ the wavefuction (dashed-dotted curve)
spreads out along the $z$-direction. Wavefunctions for $m>m_{cr}$ (not shown in the
figure) are localized on the two boundaries $z=\pm 1/{\alpha}$. 

The spectrum is continuous in both cases and these are not bound
states, in contrast to the case of \cite{randal99} for the
massless graviton mode. This is easily seen from the form of the
corresponding zero energy eigenvalues in the Schr\"odinger
equation (\ref{theequation}). However {\it there is localization}
within the horizon of a {\it thin band} of Kaluza-Klein modes,
with masses up to $\sqrt{2}\alpha$.

This leads to modifications of the Newton's law on the brane hypersurface
$\{ X^I \}, I=0, \dots 3$ at $z=0$ (where the light band of modes
including the massless one peaks, and where the cosmological
constant vanishes). This can be easily calculated from the
corresponding Green's function in the static potential between two
mass sources $m_1,m_2$. The corrections are generated by the
exchange of massive modes with masses up to $\sqrt{2}\alpha$:
\begin{eqnarray}
&~& V(r) \sim G_N \frac{m_1 m_2}{r} \left( 1 + \int _0 ^{\sqrt{2}\alpha}
\frac{1}{M_s^2}me^{-mr}dm \right) = \nonumber \\
&~& G_N \frac{m_1 m_2}{r} + \frac{G_N m_1 m_2}{r^3 M_s^2}\left(1 - e^{-\sqrt{2}\alpha r}\right)
\simeq G_N \frac{m_1 m_2}{r} + \frac{G_N m_1 m_2 \sqrt{2}\alpha }{r^2 M_s^2}
\label{newton}
\end{eqnarray}
for small thickness of the band $\alpha$. Thus, we see that the
presence of massive Kaluza-Klein modes in the space-time
(\ref{bulk}), due to the presence of a recoiling brane (our
world), struck by another string soliton, results in attractive
corrections to Newton's law of $r^{-2}$ scaling, which are
suppressed by a power of $\alpha/M_s$. In our case above the
string scale may be assumed to be the conventional one, unlike the
models of \cite{dimopo98,antoniadis99}, $M_s \sim 10^{18}$ GeV. In
the (formal) limit $\alpha \rightarrow \infty$ one recovers the
situation discussed in \cite{randal99}, but as we have explained
above our logarithmic formalism is valid for small $\alpha$.

The existence of thin bands of Kaluza-Klein modes sets a new mass hierarchy,
in the sense that masses of order $\alpha$ may determine
the supersymmetry-breaking scale on the brane $X^I$, due to the recoil
process, as discussed in \cite{adrian+mavro99}, where we refer the
reader for details. Here we simply mention that
the fact that the vacuum energy $V_{(z=0)}$ on the initial brane located at
$z=0$ is positive and of order $12\alpha^2$ (c.f. (\ref{explsol}))
signals supersymmetry breaking on the brane at a scale determined by
$\alpha << M_s$, which may be taken to
be of TeV scale, according to standard arguments on the gauge hierarchy
problem of quadratic divergencies in
four-dimensional
spontaneously broken gauge theories.
Notice that our picture for the hierarchy
is different from that of \cite{randal99}, where the low mass scale
(compared to $M_s$) on our world arises
because of the small overlap of the
bulk graviton wavefunction with that on the brane.
In our picture, such small scale factors are valid only on the horizon at
$|z|=1/\alpha$, which, however, lies
far away from the observable world.

The above considerations concern one non-compact bulk direction.
The issue of compact bulk directions is complicated in our case
because of the form of the metric (\ref{bendinglineel}), which
implies deficits (c.f. (\ref{deficit})),
as discussed in \cite{emw99}. It will be left
for a future publication. Nevertheless, we believe that the
results presented here are
sufficient to demonstrate the important r\^ole of
recoil (and in general quantum fluctuations) on the physics of
large extra dimensions in string theory. The formation of bands of
Kaluza-Klein modes, localized within the horizon of the metric
(\ref{bulk}) occurs here without the necessity of considering
periodic lattice of branes. And actually the localization is
obtained in a dynamical way, consistent with conformal field
theory on the world-sheet of the underlying string theory.

\section{Conclusions}

In this paper, we have made an attempt to generate  dynamically
the $M_{Planck}-M_W$ scale hierarchy in the context of $D$-brane
scattering.  Assuming a $D4$-brane embedded in a five dimensional
space-time  we showed that scattering with another $D$-brane
generates a bulk AdS5 space-time. The original $D4$-brane
located at the origin of the fifth dimension,
is interpeted as our Euclideanized four-dimensional
space-time,
where  the cosmological constant is found to be  zero.
Taking into account deformations due to incorporated
recoil effects,  we calculated the space-time metric and showed that
it satisfies the classical  Einstein equations. Solving the
linearized equation for the graviton modes, we find that
there appears a band of massive lower Kaluza-Klein excitations,
including the massless ordinary graviton state, which is localized
in a small region of the fifth dimension around the origin where
the $D4$-brane is located.

More precisely, due to 
the recoil quantum fluctuations of the $D4$-brane there
is localization of a (continuous) thin band
of massive KK states
with masses up to $\sqrt{2} \alpha$,
for small $\alpha << M_s$,
where the parameter $\alpha$ is related to the strength of the
quantum fluctuations and sets the supersymmetry-breaking
scale on the $D4$-brane. In this sense, the above
approach generates dynamically a mass hierarchy,
given the smallness of $\alpha$ as compared to the string scale
$M_s$, which in our approach is assumed of order $10^{18}$ GeV.
There is also the appearance of an horizon
located at $|z|\sim 1/\alpha$.

At present, our considerations pertain to non compact fifth
dimension. In the case of compact bulk dimensions there is a discrete
set of allowed KK states, with masses quantized in units of the
radius of the compact dimension. However, in that case,
within the context of the
recoil approach, there are
induced deficits~\cite{emw99} which complicate the analysis.
Such issues, togheter with the
extension of the above approach
to include more than one extra bulk dimension, are left for future work.
However, we believe that the results presented here are of sufficient
interest to motivate further studies along this direction.

\section*{Acknowledgements}

The work of N.E.M. is partially supported by P.P.A.R.C. (U.K.)
through an advanced fellowship.

\end{document}